\documentclass[prb,twocolumn,superscriptaddress]{revtex4-2}

\usepackage{amsmath,amssymb,mathrsfs}
\usepackage{graphicx}
\usepackage{colortbl}
\usepackage{ulem}
\usepackage{CJK}

\begin{document}
\title{Nearly degenerate ground states of a checkerboard antiferromagnet and their bosonic interpretation}
\author{Haiyuan Zou}  
\altaffiliation{hyzou@phy.ecnu.edu.cn}
\affiliation{Key Laboratory of Polar Materials and Devices (MOE), School of Physics and Electronic Science, East China Normal University, Shanghai 200241, China}

\author{Fan Yang} 
\affiliation{Department of Physics, Beijing Institute of Technology, Beijing 100081, China}

\author{Wei Ku} 
\altaffiliation{weiku@sjtu.edu.cn}
\affiliation{Tsung-Dao Lee Institute, Shanghai 200240, China}
\affiliation{Key Laboratory of Artificial Structures and Quantum Control (Ministry of Education), Shanghai 200240, China}
\affiliation{Shanghai Branch, Hefei National Laboratory, Shanghai 201315, China}

\begin{abstract}
The spin-$1/2$ model system with antiferromagnetic (AF) couplings on a $J_1$-$J_2$ checkerboard lattice, known as the planar pyrochlore model, is strongly frustrated and associated with a two-to-one dimensional crossover. 
Using the Projected Entangled Simplex States tensor network ansatz, we identify a large number of nearly degenerate states in the frustrated region ($J_1<J_2$).
Specifically, we find the long-sought crossed-dimer valence bond solid (VBS) state to be the ground state at $J_1\lesssim J_2$, while various 1D AF correlated states take over the rest.
We verify the stability of the VBS state against nematic perturbation.
The corresponding bosonic picture provides an intuitive understanding of the low-energy physics.
Particularly, it predicts weaker VBS states in the easy-plane limit, which we confirm numerically.
Our results clarify the most essential ground state properties of this interesting system and demonstrate the usefulness of bosonic picture in dealing with frustrated magnetism.
\vspace{0.2cm}
\\\textbf{tensor networks, frustrated magnetism, dimensional crossover, degenerate states}
\\\textbf{PACS number(s):} 05.10.Cc, 64.60.Cn, 75.10.Jm
\end{abstract}

\maketitle
\section{Introduction}
Determining the many-body ground states and classifying the phase transitions in strongly correlated systems are two central quests of condensed matter physics. For these purposes, physicists are interested in the low-dimensional frustrated systems which defeat long-range orders and exhibit exotic paramagnetic ground states, such as the quantum valence bond solid (VBS)~\cite{ReadSachdevVBS,DSWangPRL2018,SO5PRL2015,DeconfinePRL2013} and the spin liquid~\cite{ANDERSON1987,Savary2016,RMP2017SL}. Meanwhile, the past few decades witnessed the fast development of quantum fluctuation driven phase transition~\cite{QPTsachdev} beyond the conventional Landau's symmetry-broken mechanism, such as transitions characterized by deconfined quantum critical points~\cite{Deconfined2004} and symmetry fractionalization~\cite{SPT}. Exploring the
complex emergence behaviors in frustrated systems, which go beyond a small number of universal set of rules,
might give valuable insights into the understanding of the high temperature superconductivity~\cite{RMP06Doping} and the realization of quantum computing~\cite{Kitaev20032}. Realistically, frustration is induced by the combined effects from nontrivial lattice geometries~\cite{diep2013}, competition of couplings~\cite{J1J2_1}, and exotic interactions~\cite{Kitaev20062}.    
Although great theoretical progress and fast development of many techniques have been made, the nature of the ground state of many frustrated systems still remains controversial~\cite{Yan2011,Gapless2013,Liao2017Gapless,Jiang2012J1J2,J1J2Plaq2014,Wang2016J1J2}. It is important to consider these problems from a different angle.

\begin{figure}[t]
\hspace{-0.3cm}
\includegraphics[width=0.4\textwidth]{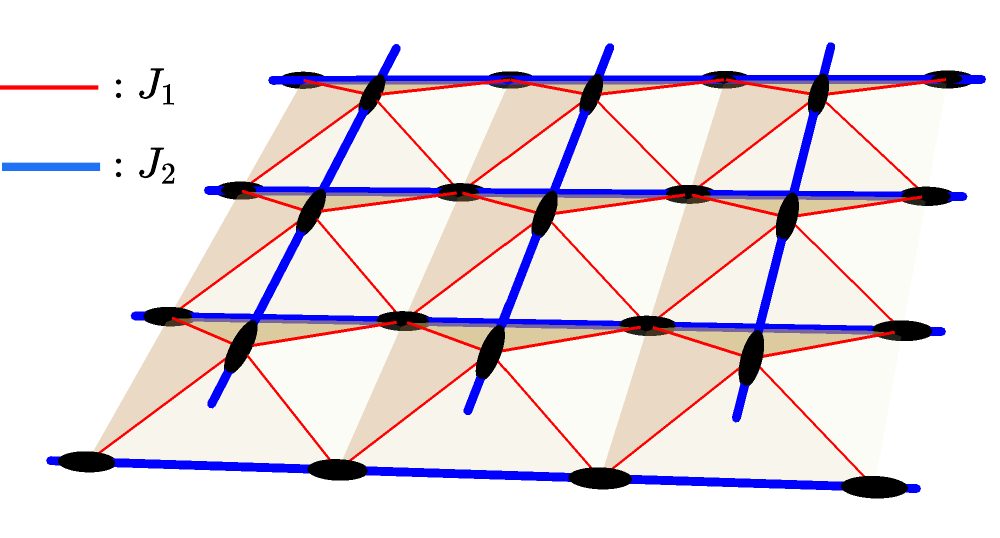}
\caption{Planar pyrochlore model in the 1D dominant region.  Lattice sites (ovals) are coupled along the 1D chain via the stronger $J_2$ (blue) bonds, followed by weaker 2D coupling along the $J_1$ (red) bonds.}
\label{fig:model}
\end{figure}

The antiferromagnetic (AF) spin-$1/2$ Heisenberg model on a two-dimensional (2D) checkerboard lattice, also known as the planar pyrochlore model, is a paradigmatic example with highly frustrated interactions.
It is described by the Hamiltonian:
\begin{equation}
H=J_1\sum_{\rm NN}\mathbf{S_i}\cdot\mathbf{S_j}+J_2\sum_{\rm NNN}\mathbf{S_i}\cdot\mathbf{S_k},
\label{eq:model}
\end{equation}
where $\mathbf{S_i}$ is the spin operator on site $i$, $J_{1,2}$ are the exchange couplings between the spins with the nearest neighbors (NN) and the next nearest neighbors (NNN), respectively. Note that in the easy-plane limit ($XY$-limit) of the model, each bond energy $J_{1,2}\mathbf{S_i}\cdot\mathbf{S_j}$ in the Hamiltonian with the Heisenberg limit is simply replaced by $J_{1,2}(S^x_iS^x_j+S^y_iS^y_j)$. At $J_1\gg J_2$, the system has a ground state with 2D AF correlation along $J_1$ bonds (cf: red bands in Fig.~\ref{fig:model}), leaving a strongly ferromagnetic correlation along the $J_2$ (blue) bonds.
At the opposite limit $J_2\gg J_1$, the system reduces to one-dimensional (1D) AF correlated chains along the $J_2$ (blue) bonds weakly coupled by geometric frustrated $J_1$ interaction, drastically incompatible with the 2D limit.
The corresponding ground state is believed to be a sliding Luttinger liquid~\cite{Liquid2002}.
Thus, the system hosts a two-to-one dimensional (2D-to-1D) crossover reflecting the severe competition between these two interactions.

This frustrated system with the dimensional crossover feature has attracted tremendous studies using different techniques~\cite{Liquid2002,StrongC2002,PVBSED2003,CORE2003,CDM2005,PEPS2011,DMRG2008,CCM2012}. In the region where $J_2\sim J_1$, it is believed that the ground state is a plaquette-VBS (P-VBS) state~\cite{StrongC2002,PVBSED2003,CORE2003}. 
However, the true ground state at $J_2>J_1$ is still under serious debate.
An earlier proposal~\cite{CDM2005} suggested a crossed-dimer type VBS (CD-VBS) state, which was supported by results from the Density Matrix Renormalization Group~\cite{DMRG2008} and the coupled cluster method~\cite{CCM2012}.
On the other hand, tensor networks with Projected Entangled Pair States ansatz found a stripe 
\begin{widetext}

\begin{figure}[t]
\includegraphics[width=1\textwidth]{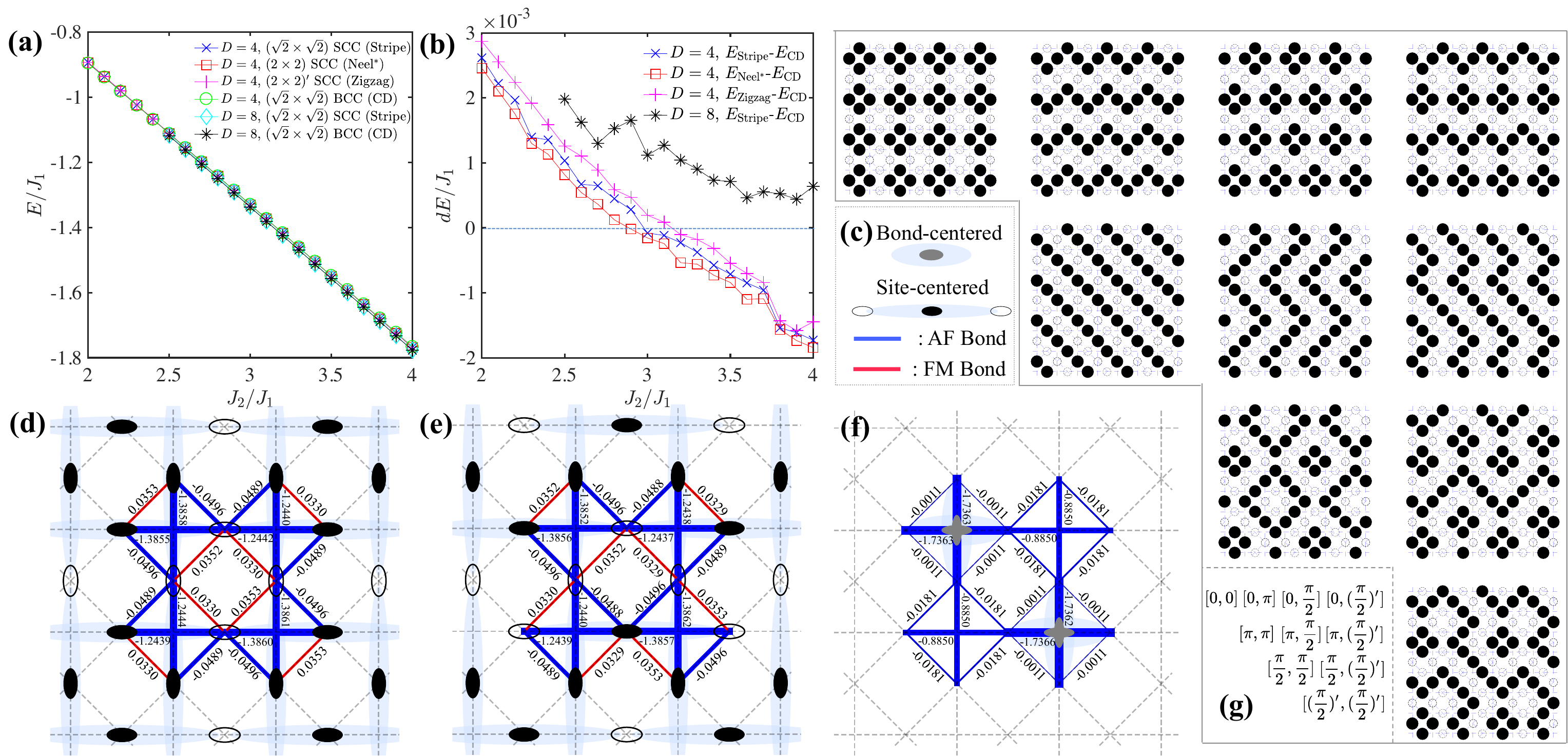}
\caption{(a) Energies of nearly degenerate states as functions of $J_2>J_1$ with $D=4$ and $D=8$. (b) The same using the CD (BCC) state as a reference, showing that this degeneracy is enhanced, and the level crossing between the CD (BCC) and SCC states appears at larger $J_2/J_1$ as $D$ increases. (c) Illustration of the spread of the bosonic orbitals in BCC and SCC states in panels (d-f), in which blue and red lines demote AF and FM bonds. (d-f) Examples of the bond energies in various nearly degenerate states corresponding to different bosonic charge correlations at $J_2/J_1=3$ in a $2\times 2$ PESS unit cell: (d) $(2\times 2)$ SCC (N\'eel*), (e) $(2\times 2)'$ SCC (zigzag), and (f) $(\sqrt{2}\times\sqrt{2})$ BCC (crossed-dimer). (g) All 1D AF correlated states found in a larger $(4\times 4)$ unit cell. The label with $[x,y]$, ($x=0,\pi,\pi/2,(\pi/2)'$) is described in the text.}
\label{fig:conf}
\end{figure}

\end{widetext}
ordered phase instead~\cite{PEPS2011}. In addition, the nature of the phase transitions between these different possible ground states is still unclear~\cite{CDM2005}.

One should, however, be cautious about detailed results from the tensor network methods~\cite{PEPS1,iPEPS,TEBD,Schollwck2011,Ors2014,MERA,TRG,TERG,PESS,DMRG}, as the dominant quantum entanglement of the states is directly affected by the particular ansatz employed in the implementation, in addition to the control parameter ``virtual bond dimension''.
Consequently, the best choice of ansatz is highly dependent on the specific geometrical structure of the system~\cite{Corboz2012,PESS}.

Here, using a recent developed tensor network ansatz, i.e., the projected entangled simplex states (PESS)~\cite{PESS} in the thermodynamic limit, we obtain a large number of nearly degenerate ground states in the $J_1 < J_2$ region, including the long-sought CD-VBS state.
The existence of these highly degenerate states explains why the ground state is elusive in previous studies and demonstrates the superiority of PESS in capturing both 1D and 2D entanglement of this system in the 2D-to-1D crossover region~\cite{PESS,noteS}.
Specifically, while these 1D AF correlated states are slightly lower in energy for $J_1\ll J_2$, the CD-VBS state becomes the ground state at $J_1 \lesssim J_2$.
We further confirm the stability of the CD-VBS state by exposing it to a nematic perturbation.
An intuitive understanding for these states emerges through different local charge correlations in a bosonic picture, which helps elucidate the nature of the phase transitions.
The picture also points out a slight disadvantage of the 1D AF correlated states due to the effective repulsion.
We verify this analysis by showing a smaller CD-VBS region in the $XY$-limit of the model.
Interestingly, at the $J_1 \gtrsim J_2$ we not only find the known P-VBS state but also find a surprising out-of-plane spin-configuration in the $XY$-limit, corresponding to a bosonic charge-density-wave (CDW).
Our results solve the ground state question of this fascinating system, and exhibit a useful bosonic picture for frustrated magnetism.

\section{Method and Results}
We first use a $2\times 2$ (with a $J_2$ bond as one lattice spacing) PESS~\cite{PESS} unit cell which contains eight different local tensors, to explore the checkerboard AF model (Eq.~\ref{eq:model}).
The size of the unit cell (i.e., the size of the translational unit in spontaneously symmetry broken states) turns out to be an important ingredient in our thermodynamic limit calculation. Starting from random states, the imaginary time evolution update method evolves the initial states to different converged final states. We find that a ``virtual bond dimension'' $D=4$ already
reach lower energies than other methods and our results are further confirmed by larger $D$, larger unit cell, and full update calculations~\cite{noteS}. 

Figure.~\ref{fig:conf}(a) shows nearly identical energy for different states at $J_2>J_1$ found in our calculation.
Particularly, we find direct evidence of a CD-VBS state~\cite{CDM2005} [Fig.~\ref{fig:conf}(f)], in addition to the stripe state and N\'eel* state [Fig.~\ref{fig:conf}(d)]. Note that the conventional name ``stripe" and ``N\'eel*" here are not local magnetic orders but correspond to different 2D patterns resulting from the interplay among various 1D AF correlated chains~\cite{noteS}.
Surprisingly, a new zigzag correlation with a $2\times 2$ period [labeled as $(2\times 2)'$ in [Fig.~\ref{fig:conf}(e)] and another seven similar states with a larger period [Fig.~\ref{fig:conf}(g)] appear in the $4\times 4$ unit cell calculation~\cite{noteS}.
All states in (g) are similar 1D AF correlated states different from the CD-VBS state due to qualitatively distinct entanglements.

To better understand these states, we find it is very helpful to employ a bosonic picture.
Using the fully down-spin polarized state as a reference, the local spin state can be represented by a hard-core boson, $b^\dagger$ with $b^\dagger_ib^\dagger_i=0$, through $S^+_i\rightarrow b^\dagger_i$, $S^-_i\rightarrow b_i$, and $S^z_i\rightarrow n_i-1/2$ ($n_i\equiv b^\dagger_ib_i$).
Table~\ref{tb:spvsb} translates the most relevant physics of our spin system into the bosonic picture.
Intuitively, the AF Ising coupling $S^z_iS^z_j$ becomes an effective repulsion between two bosons or two holes, disfavoring the parallel alignment of neighboring spins.
Most importantly, the spin-fluctuation $S^x_iS^x_j+S^y_iS^y_j$ conveniently becomes the kinetic energy of a boson, and correspondingly, the two-body spin-singlet state is merely a single-particle anti-bonding bosonic state.
Such a reduction from a two-body physics to a single-particle one provides a significant simplification in systems with strong fluctuations in general.
\begin{table}[th]
\begin{tabular}{p{2cm}p{2cm}p{1cm}p{2cm}p{2cm}}
\cline{1-5} 
\multicolumn{2}{l}{Spin picture} & &  \multicolumn{2}{l}{Bosonic picture}\\
\cline{1-5} 
\noalign{{\color{white}\hrule height 1pt}}
\multicolumn{2}{l}{$S^x_iS^x_j+S^y_iS^y_j$}& &\multicolumn{2}{l}{$b^\dagger_ib_j+b^\dagger_jb_i$ (kinetic energy)}\\
\noalign{{\color{white}\hrule height 1pt}}
\multicolumn{2}{l}{$S^z_iS^z_j$}& & \multicolumn{2}{l}{$(n_i-1/2)(n_j-1/2)$ (repulsion)}\\
\noalign{{\color{white}\hrule height 1pt}}
\multicolumn{2}{l}{$\frac{1}{\sqrt{2}}(|\uparrow\downarrow\rangle-|\downarrow\uparrow\rangle)$}& &\multicolumn{2}{l}{$\frac{1}{\sqrt{2}}(|10\rangle-|01\rangle)$ (anti-bonding)}\\
\cline{1-5}
\end{tabular}
\caption{\label{tb:spvsb} The spin and boson correspondence.}
\end{table}

For example, in our AF checkerboard lattice, the lack of global magnetization implies that our bosonic system should generally be half-filled, making them easily stuck into some sort of local charge order.
The larger $J_2b^\dagger_ib_j$ dictates an alternating charge order along the $J_2$ direction for low-energy states, such that each boson has a chance to extend its orbital to lower its kinetic energy.
Next, the smaller $J_1b^\dagger_ib_j$ would similarly prefers to leak into as many empty $J_1$-neighbors as possible.

\begin{figure}[h]
\includegraphics[width=0.48\textwidth]{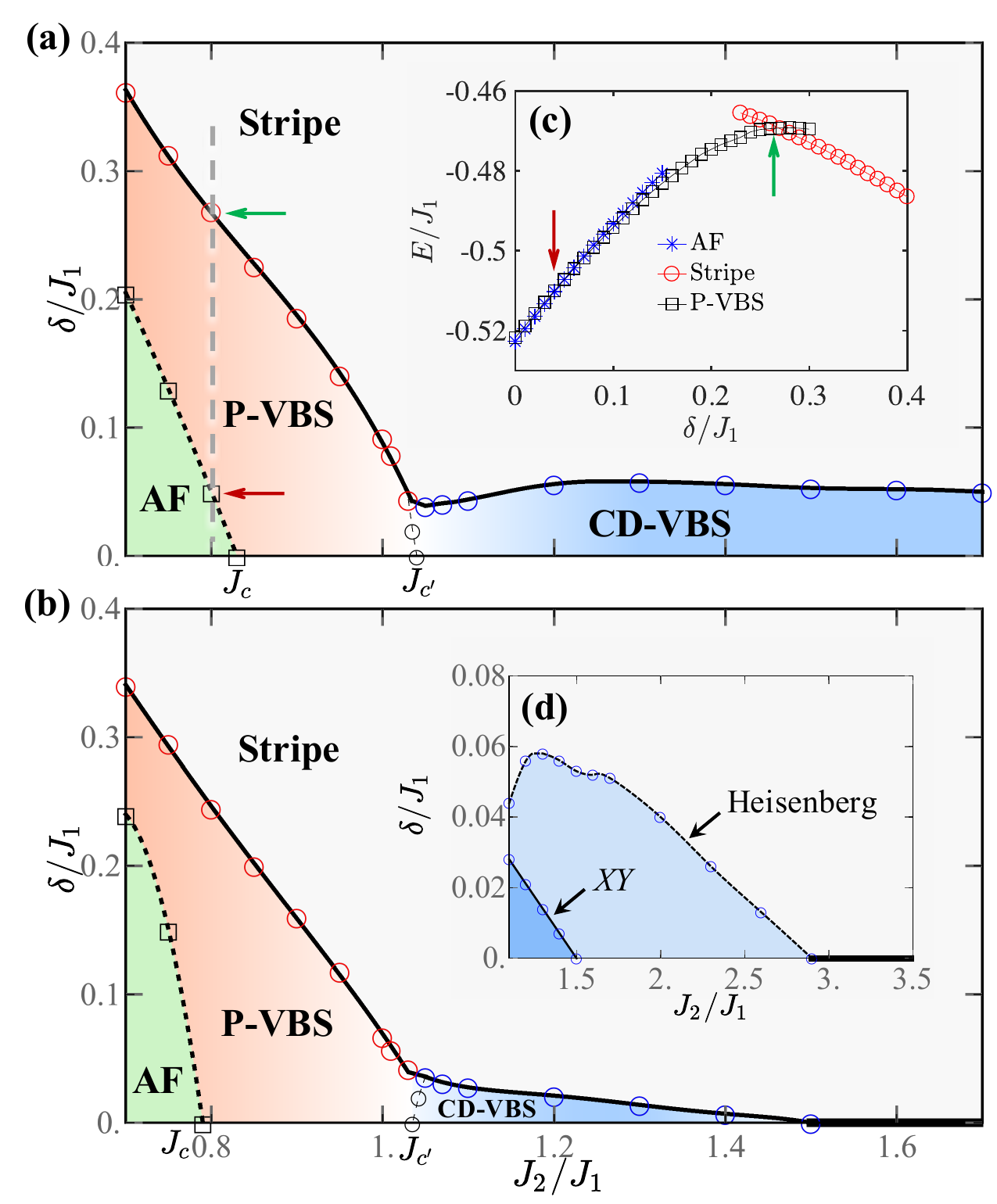}
\caption{$\delta$-$J_2$ phase diagrams of the model in (a) the Heisenberg limit and (b) the $XY$-limit demonstrate AF (green), P-VBS (red), CD-VBS (blue), and stripe (grey) phases. 
Panel (c) illustrates the energies of the most competitive states as a function of $\delta$ along the grey long dashed line at $J_2=0.8J_1$, indicating strong (green arrow) and weak (red arrow) first-order transitions.
Correspondingly, the solid and dotted lines in (a)\&(b) denote phase boundaries of strong and weak first-order transitions, respectively.
Panel (d) shows that in the $XY$-limit, the CD-VBS phase starts at a lower value of $J_2/J_1$, above which multiple nearly degenerate SCC states take over, denoted by the thick solid lines.} 
\label{fig:phaseD}
\end{figure}

We now demonstrate that this framework provides a very intuitive understanding of the low-energy states of the systems.
First, it is easy to understand the nearly identical energies of the 1D AF correlated states in Fig.~\ref{fig:conf}(g).
These states correspond to various site-centered charge correlated (SCC) bosonic states, in which each boson is surrounded by two unoccupied sites along the $J_2$ and $J_1$ directions, allowing its orbital to extend and in turn lowering its kinetic energy.
With finite unit cells, we find all the possible SCC states, i.e., the stripe states with a $\sqrt{2}\times\sqrt{2}$ period, the additional ``N\'eel*" and ``Zigzag" states with a $2\times 2$ period, and seven other states with a $4\times 4$ period. The bosonic picture provides a systematic way to classify all these newly found states. The repeating charge patterns of the vertical and horizontal 1D AF correlated chains along $J_2$ directions give different nesting vector $[x,y]$ for these states. For example, the ``zigzag" pattern, with the vertical bosons repeated for every lattice space while the horizontal ones are repeated every other lattice space, is labeled as $[0,\pi]$. Other states classified in this notation are shown in Fig.~\ref{fig:conf}(g).
Similarly, the long-sought CD-VBS state also corresponds to a charge correlated state, except that each boson now resides in a bond-centered anti-bonding orbital.
It turns out that the CD-VBS state is the only bond-centered charge correlated (BCC) state with energy that can be quite similar to those SCC states. Other cases are higher energy states with blocking of $J_1$ orbitals~\cite{noteS}.
We anticipate the usefulness of such a bosonic picture to extend to other frustrated spin systems with nearly degenerate states~\cite{WhiteHoneycomb2013}.

The near degeneracy between the CD-VBS and SCC states is not expected to persist in the $J_1 \rightarrow 0$ limit, where the pure 1D solution is exactly known~\cite{Bethe1931} to have only weak dimerization in terms of the bond energy, similar to those in the SCC states.
This is not too surprising as these two sets of states are quite different in their bosonic orbital structure discussed above, with SCC having a larger extent of the orbital along the $J_2$ directions, as shown in Fig.~\ref{fig:conf}(c).
Indeed, Fig.~\ref{fig:conf}(b) shows that the energy of the CD-VBS state approaches the latter in the vicinity of $J_2\sim 3J_1$ in the $D=4$ PESS calculation. This behavior is qualitatively justified in larger $D$ calculations, which show that the competition
between CD-VBS and SCC states becomes even stronger and gives a closer energy level crossing at larger $J_2/J_1$~\cite{noteS}. 
In fact, the CD-VBS state becomes the ground state near $J_1 \lesssim J_2$ due to its avoidance of the energy cost associated with the frustrated $J_1$ bond (blocked bosonic kinetic process) present in the 1D AF states [c.f.: red bonds in Fig.~\ref{fig:conf}(d)(e)].

One can examine the stability of the CD-VBS state by lowering the energy of the stripy SCC state in Fig.~\ref{fig:conf}(e) via the introduction of NN bond anisotropy $\delta=J_1-J_1'>0$, with the directions of $J_1'$ and $J_1$ perpendicular to each other.
Figure~\ref{fig:phaseD}(a)\&(d) give the resulting phase diagram with a large region of CD-VBS state under weak anisotropy.
The finite region of the VBS state confirms that it is indeed a stable ground state at $J_2/J_1 \lesssim 3$ in our calculation.

Upon careful examination of these nearly degenerate states, we found that SCC states suffer more from the effective $J_1$ repulsion, as demonstrated by the red bonds in Fig.~\ref{fig:conf}(d)(e).
In contrast, owing to its bond-centered nature, the CD-VBS state experiences approximately an equal amount of repulsion and attraction, resulting in a nearly decoupled crossed dimer [c.f.: the thinnest blue bond in Fig.~\ref{fig:conf}(f)].
Therefore, one should be able to further strengthen the SCC phase by relieving the repulsion, via the removal of the $S^z_iS^z_j$ coupling (i.e., the $XY$-limit).

Indeed, our resulting phase diagram of the $XY$-limit calculation shown in Fig.~\ref{fig:phaseD}(b)\&(d) confirms this intuition.
The energy of a SCC state is now more competitive, such that the CD-VBS phase shrinks to a much smaller region $J_2/J_1\lesssim 1.5$.
Correspondingly, the amount of anisotropy needed to destroy the CD-VBS phase is also smaller in this case.

\section{Discussion}
Our results above can provide useful insights to some of the real materials.
For example, the pyrochlore material GeCu$_2$O$_4$~\cite{Yamada2000,uudd2016} is estimated to have $J_2/J_1\sim 6$~\cite{Yamada2000}, well within the nearly degenerate region. Experimentally, the true ground state of GeCu$_2$O$_4$ is still under debate, as both AF correlation~\cite{Yamada2000} and up-up-down-down (uudd) correlation~\cite{uudd2016} are detected.
Using the string correlation parameter~\cite{noteS,string1989,ZouSPT2019,ZouSPT2020} as a measure of the uudd correlation, we find it negligible in the SCC states but rather strong in the CD-VBS state, with a slow decay spanning over more than 400 sites~\cite{noteS}.
Generally the near degeneracy implies that the system is highly susceptible to various weak perturbations in real materials~\cite{Yamada2000,uudd2016,ZnVO04}, and it might even induce spontaneous symmetry breaking that lowers the degree of degeneracy.

\begin{figure}[h]
\includegraphics[width=0.45\textwidth]{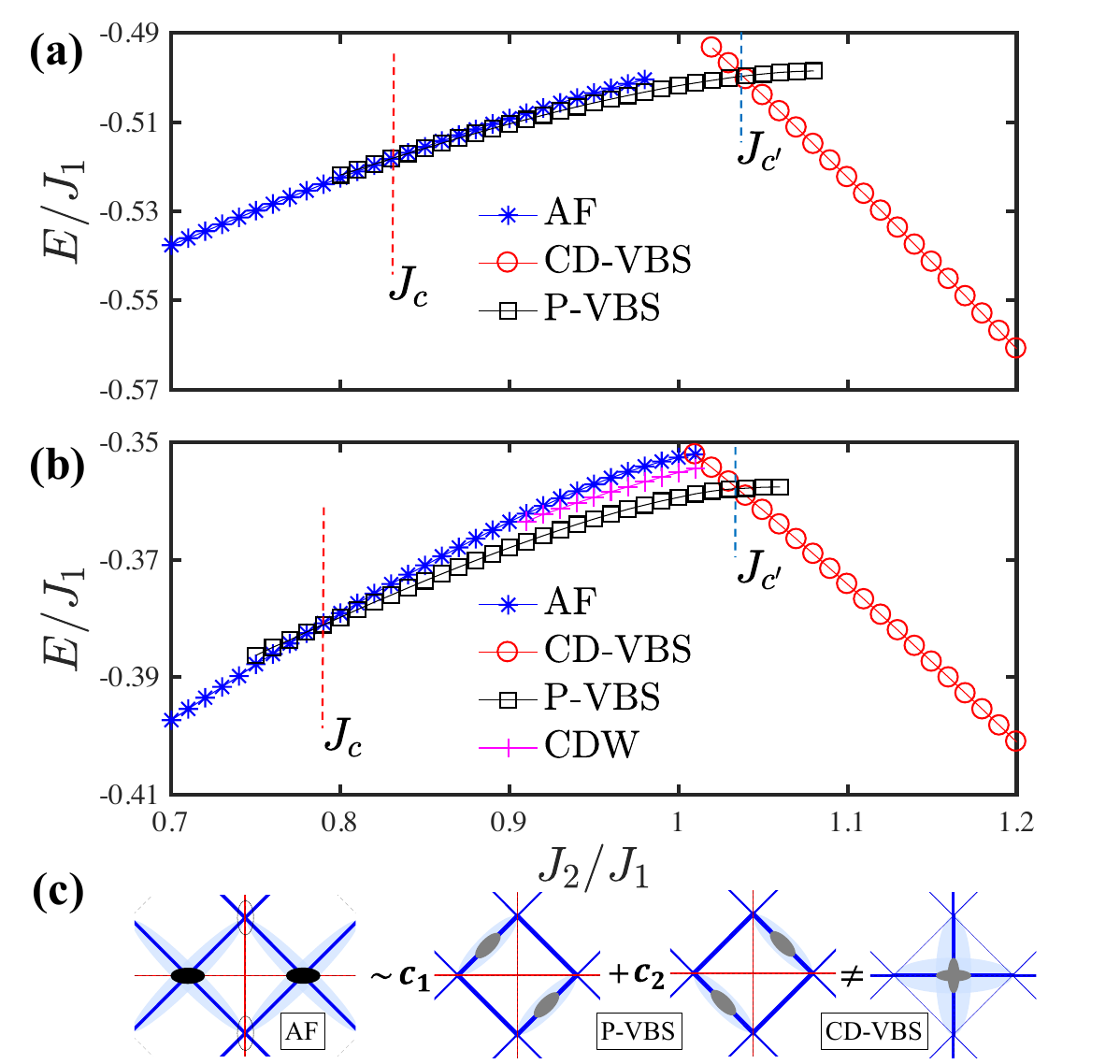}
\caption{$J_2$-dependent energies of competing states near the quantum phase transitions in the (a) Heisenberg limit and (b) $XY$-limit for $\delta=0$ show a weak first-order phase transition at $J_c$, given the similar bosonic orbital structures in (c).
In contrast, the transition at $J_{c^\prime}$ is strongly first-order, owing to the very different orbital structures in (c) corresponding to the different dominant couplings along $J_1$ and $J_2$ bonds.
Panel (b) also shows the existence of a meta-stable CDW state containing only out-of-plane spin components.
}
\label{fig:transition}
\end{figure}

Generally, the emergence of VBS like states appears to be a generic feature in the strongly frustrated region in many models.
Our above bosonic analysis on the emergence of BCC-type VBS state points out a natural mechanism to explain such a rather generic trend.
Typically, AF spin systems correspond to half-filled hard-core bosonic system, which easily get jammed into a local charge correlated structure.
In the absence of strong frustration, the larger size of site-centered orbitals would benefit more from the kinetic energy and thus tend to be the dominant ground states.
On the other hand, in the presence of strong frustration, bond-centered orbitals, having more space to maneuver collectively, are able to partially avoid the energy cost of the frustration and thus can eventually take over.

In the 2D checkerboard lattice, another VBS state named P-VBS is known to appear in the 2D dominant ($J_1\gtrsim J_2$) region (Fig.~\ref{fig:phaseD} and Fig.~\ref{fig:transition}).
Figure~\ref{fig:phaseD} shows that our calculation reproduces this P-VBS state between the 2D AF phase and the stripe phase.
Furthermore, Fig.~\ref{fig:transition}(c) demonstrates that the formation of the P-VBS state can be understood via the above mechanism as well.
In the weakly frustrated region, the ground state is the 2D AF state corresponding to a SCC state with alternating charge along the $J_1$ directions.
As $J_2$ grows, this state will suffer more and more from the blockage of $J_2$ paths. 
Eventually at $J_2/J_1=J_c\sim 0.8$, the P-VBS will emerge since it is composed of a superposition of $J_1$-bond-centered orbitals that allow potential kinetic processes toward the $J_2$ directions.
This confirms the applicability of the above general picture.

Similar to the 1D dominant region, the SCC 2D AF state has a similar energy to the BCC P-VBS state, as shown in Fig.~\ref{fig:transition}.
Thus, we anticipate a weak first-order transition at $J_c\sim 0.8$.
This result is in agreement with a previous conclusion~\cite{CDM2005} that a deconfined phase transition is highly unlike in this system.
In contrast, the change from the 2D dominant region ($J_2/J_1 < 1$) to the 1D dominant one  ($J_2/J_1 > 1$) involves bond-centered orbitals containing different bonds.
One thus expects a strong first-order phase transition at $J_{c^\prime}\sim 1.04$, supported by the sharp contrast in energy in Fig.~\ref{fig:transition}.

As a side note, in the $XY$-limit between $J_c$ and $J_{c^\prime}$ [Fig.~\ref{fig:transition}(b)], we find a unexpected excited state containing an AF correlation that is completely out-of-plane.
This state can be understood as a CDW state in the bosonic picture due to enhanced site-centered charge correlation, in associate with the frustration induced suppression of kinetic effects that couple symmetry related degenerate states~\cite{noteS}.

\section{conclusions}
In summary, using PESS, a novel tensor network method, we demonstrate a large number of nearly degenerate states in the checkerboard AF system, related to the lower dimensional decoupling featured in this model, which implies additional phase transitions.
Particularly in the 1D-dominant frustrated region, our calculation produces the long-sought CD-VBS state, which becomes the ground state at $J_2\gtrsim J_1$ and then loses to the 1D AF correlated states at $J_2\gg J_1$.
We verify the stability of the VBS state via the introduction of a nematic background and find it resilient against a finite strength of nematicity.
A simple understanding of our results can be provided by a bosonic picture.
The picture also suggests that the CD-VBS state benefits from the charge repulsion (the $S^zS^z$) effect, which we confirm by realizing a weaker CD-VBS state in the $XY$-limit of the model.
Our results resolve a long-standing debate in the active field of quantum frustrated magnetism and showcase the rich phenomena near quantum phase transition in strongly correlated systems.

\section{acknowledgments}
We thank Anthony Hegg, Ruizheng Huang, Wei Li, Haijun Liao, Sudeshna Sen, Ling Wang, Rui Wang, Ning Xi, Tao Xiang, and Zhiyuan Xie for helpful discussions. This work is supported by the National Natural Science Foundation of China (NSFC)~$\#$12274126.  FY acknowledges supports from NSFC~$\#$12074031 and 12234016.  WK acknowledges supports from NSFC~$\#$12274287 and 12042507 and the Innovation Program for Quantum Science and Technology~$\#$2021ZD0301900.

%

\pagebreak

\newpage

\widetext
\begin{center}
\textbf{\large Supplemental Materials for ``Nearly degenerate ground states of a checkerboard antiferromagnet and their bosonic interpretation"}
\end{center}

\setcounter{equation}{0}
\setcounter{figure}{0}
\setcounter{table}{0}
\makeatletter
\renewcommand{\thefigure}{S\arabic{figure}}
\renewcommand{\thetable}{S\arabic{table}}
\renewcommand{\theequation}{S\arabic{equation}}
\renewcommand{\bibnumfmt}[1]{[S#1]}
\renewcommand{\citenumfont}[1]{S#1}
\makeatother

\section{Projected Entangled Simplex States Tensor Networks}

\begin{figure}[h]
\centering
\includegraphics[width=0.5\textwidth]{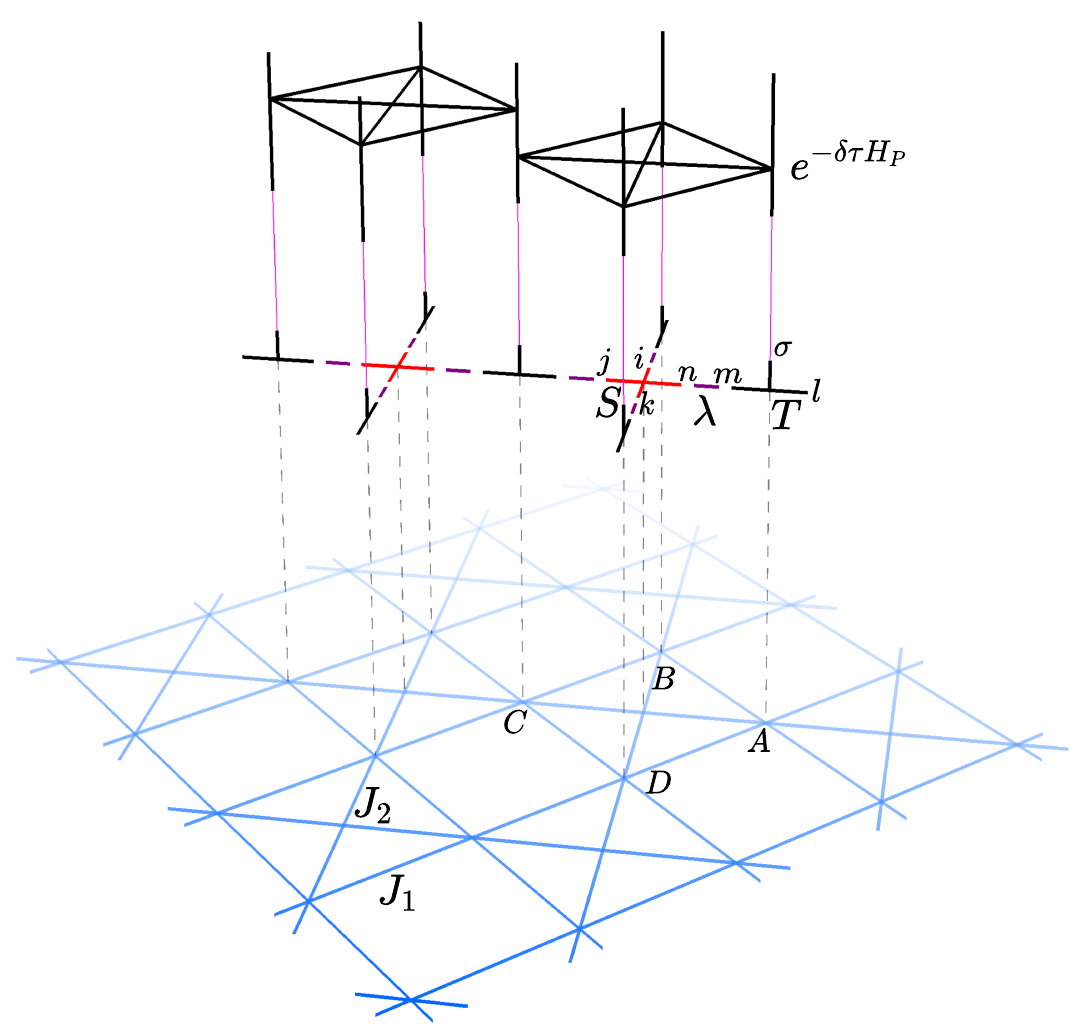}
\caption{The $J_1$-$J_2$ model with the nearest ($J_1$) and the next nearest ($J_2$) neighbor couplings on a checkerboard lattice is illustrated at the bottom. The ground state is represented by a PESS tensor network in the middle with the local tensors $T$ on the sites, the core tensors $S$ on the plaquettes with $J_2$ bonds, and the bond vector $\lambda$ in between of them. On top of the network, a symmetric updating scheme with $H_P$ on each plaquette is chosen to evolve the network to the ground state.}
\label{fig:structure}
\end{figure}

We use a recent developed tensor network ground state wave-function ansatz, i.e., the projected entangled simplex states (PESS)~\cite{PESSs}, to calculate the ground state of the many-body Hamiltonian (Eq.~1 in the main text) with $J_1$ and $J_2$ couplings on a checkerboard lattice. The basic structure of the PESS (shown in Fig.~\ref{fig:structure}) is constructed by the local tensors $T$ with physical indexes $\sigma$ on each lattice site, the simplex core tensors $S$ on the plaquettes with $J_2$ bonds, and the bond vectors $\lambda$ connecting $T$ and $S$. With all these ingredients, the ground state wave-function can be expressed as  

 \begin{equation}
|\Psi(\dots\sigma\dots)\rangle={\rm Tr}\left (\dots T^{\sigma}_{lm}\lambda_{mn} S_{nijk}\dots\right )
 \end{equation}
where $i,j,\dots ,n$ are virtual indexes characterizing the quantum entanglement of the system and Tr stands for the contraction of all the virtual indexes. 

\begin{figure}[t]
\includegraphics[width=0.7\textwidth]{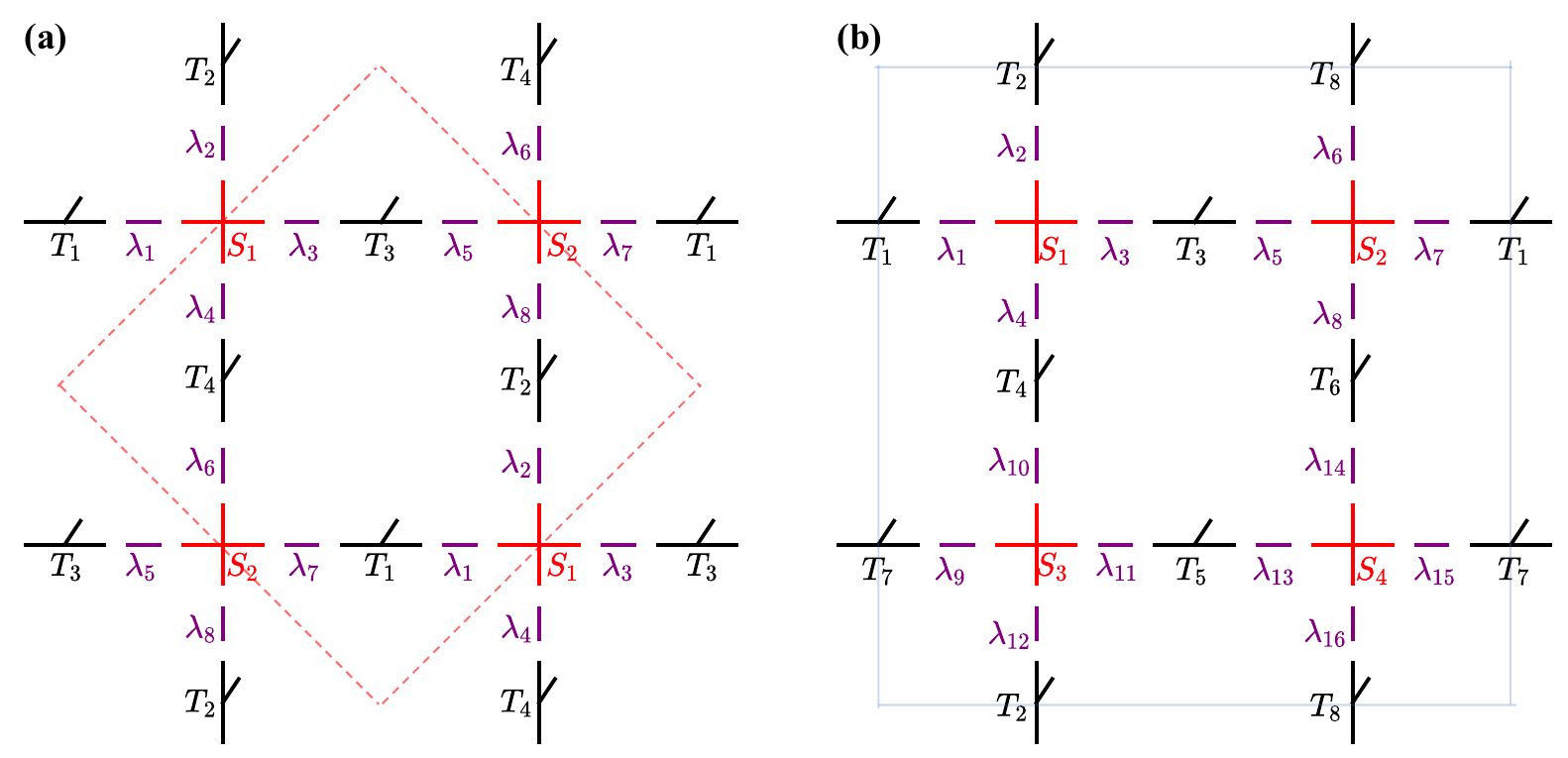}
\caption{The structure of the PESS unit cell on a checkerboard lattice. There are $4n$ local tensors $T$, $2n$ core tensors $S$, and $8n$ bond vectors $\lambda$ in one unit cell (n is an integer).  (a) Lattice constructed by the unit cell with $n=1$. The structure is repeated with a $\sqrt{2}\times\sqrt{2}$ unit cell with the dashed red boundary lines; (b) the $n=2$ unit cell with the thin blue boundary lines have a $2\times 2$ structure. (The length of a $J_2$ bond is chosen as one lattice spacing)}
\label{fig:unitc}
\end{figure}

The tensor network wave-function in the thermodynamic limit is constructed by repeating the translational invariant building blocks, i.e., unit cells in both horizontal and vertical directions in 2D. The numbers of the local ingredients $T$, $S$, and $\lambda$ in one unit cell are usually not independent due to the geometric structures. For a checkerboard lattice, it is reasonable to choose a unit cell with $4n$ local tensors $T$, $2n$ core tensors $S$, and $8n$ bond vectors $\lambda$, with n is an integer. Figure.~\ref{fig:unitc} shows two examples of the unit cells with $n=1$ and 2, with the PESS unit cell has a $\sqrt{2}\times\sqrt{2}$ structure for $n=1$ [Fig.~\ref{fig:unitc}(a)] and has a $2\times 2$ structure for $n=2$ [Fig.~\ref{fig:unitc}(b)].

Starting from an initial state $|\Psi_0\rangle$, either a random or a particular classical ordered configuration, an imaginary-time operation $\exp(-\delta\tau H)$ evolves the state to a convergent ground state as the iteration steps $N$ increases. In practice, only operation on the local tensors are performed. Trotter-Suzuki approximation is used to decompose the global Hamiltonian $H$ into local ingredient $H_P$ on each plaquette (or cluster):

 \begin{equation}
 \exp (-\delta\tau H)=\prod_P\exp(-\delta\tau H_P)+O(\delta\tau^2),
 \end{equation}
 where $H_p$ is the local Hamiltonian with spin operator $S_i$ on $i$ ($i=A,B,C,D$) (Fig.~\ref{fig:structure}) and expressed as
 \begin{equation}
 H_P=J_1(S_A\cdot S_B+S_B\cdot S_C+S_C\cdot S_D+S_D\cdot S_A)+J_2(S_A\cdot S_C+S_B\cdot S_D).
 \end{equation}
 
Using the tensor network state $|\Psi\rangle$ with convergent $T$, $S$, and $\lambda$, the physical properties, e.g., bond energy, can be calculated through 
\begin{equation}
\langle O\rangle = \langle\Psi |O|\Psi\rangle/\langle\Psi|\Psi\rangle,
\end{equation}
where $O$ is any local operator. We evaluate it using a real space coarse-graining procedure known as the corner transfer matrix renormalization group (CTMRG)~\cite{Nishino1996s,Vidal2010s} which enables one to reach the thermodynamic limit. The results are convergent with the truncation dimension of the CTMRG step larger than $D^2$. We can also use CTMRG to calculate the environment tensors for the local tensors for each time evolution step first, and then use the environment tensors to replace the bond vectors $\lambda$s to give the local tensors and the core tensors a full update (FU). FU usually gives lower ground state energy than the simple or cluster update methods.  
 
\section{Benchmark results}

We first use a $\sqrt{2}\times\sqrt{2}$ unit cell with four local tensors $T$ ($n=1$) with different virtual bond dimension $D$ form the tensors and set the time interval $\delta\tau=0.02J^{-1}$. The comparison between the PESS results (with and without FU) of the Heisenberg interaction on the checkerboard lattice and previous results from other methods are listed in Table~\ref{tb:compare} at different $J_1$ and $J_2$.

\begin{table}[th]
\begin{tabular}{|c|c|c|c|c|c|}
\hline
\hline
$(J_1,J_2)$&PESS($D=4$)&FU($D=4$)&PESS($D=6$)&PESS($D=8$)&Others\cr
\hline
$(1, 0)$&$-0.6683$&$-0.6690$&$-0.6688$&$-0.6694$&$-0.6694$ (DMRG~\cite{Stoudenmire2012s},MC~\cite{Sandvik1997s})\cr
\hline
$(1, 2)$&$-0.8956$&$-0.8965$&$-0.8998$&$-0.9024$& $-0.876$(iPEPS~\cite{PEPS2011s})\cr
\hline
\hline
\end{tabular}
\caption{\label{tb:compare} Comparison of our calculated ground state energy per site at the thermodynamics limit using PESS (both with or without FU) and the results from other methods.}
\end{table}

\begin{figure}[h]
\includegraphics[width=0.7\textwidth]{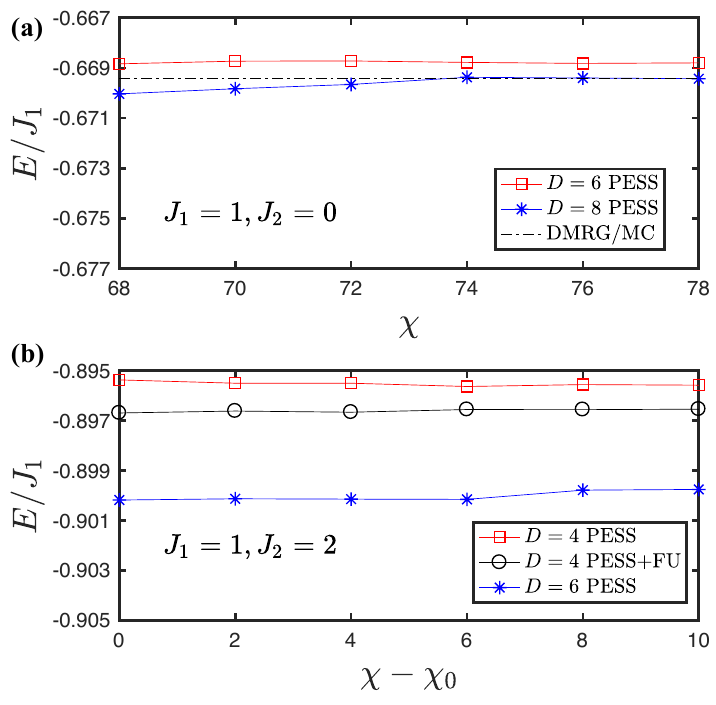}
\caption{Ground state energies with functions of truncation bond dimension $\chi$.(a) At $J_1=1,J_2=0$, energies are convergent at large $\chi$ at $D=$6,8. At $D=$8, the results are consistent with the well-known DMRG or MC results. (b) Convergent energies for $J_1=1, J_2=2$ and $D=4,6$ using PESS with and without FU at large $\chi$, where $\chi_0=D^2$.}
\label{fig:j11j22}
\end{figure}

At $J_1=1,J_2=0$, the model reduces to a 2D Heisenberg model, the ground state energy at $D=4$ in our PESS calculation is already consistent with the Density Matrix Renormalization Group (DMRG)~\cite{Stoudenmire2012s} and Monte Carlo (MC)~\cite{Sandvik1997s} extrapolated results up to $10^{-3}$, and better than the results from many other methods~\cite{RMP1991s}. Either using FU or simply increasing $D$ can improve the PESS calculation to convergent results consistent with DMRG or MC results up to $10^{-4}$. At $J_2/J_1=2$, The ground state energy of the nearly degenerate bond-centered charge correlated (BCC) and site-centered charge correlated (SCC) states obtained by PESS is lower than the energy of the stripe ordered phase from iPEPS calculation~\cite{PEPS2011s}. Either using FU or increasing $D$ gives lower ground state energies. At $J_1=0, J_2=1$, the checkerboard model is decoupled into independent 1D chains in two different directions. The results from PESS calculation are exactly the same with those from the infinite time-evolving block decimation (iTEBD)~\cite{TEBDs} calculation with the same $D$, i.e., the average energies are $-0.4410$, $-0.4425$, and $-0.4427$ for $D=4,6$, and 8, respectively, which is already consistent with the exact result $-\ln(2)+1/4$. the equivalence between the PESS and the iTEBD is due to the reason that the simplex tensor $S$ recover the geometric structure of the checkerboard lattice. At $J_1=0$, the simplex tensor $S$ is a direct tensor product of two iTEBD Schmidt matrices $\lambda$ of the two independent directions:
\begin{equation}
S'_{ij,kl}=\lambda_{i,k}\times\lambda_{j,l}
\end{equation}
where $S'_{ij,kl}$ is the reshaped matrix form of the simplex tensor $S_{ijkl}$. Thus the simplex tensor can be constructed by two independent Schmidt matrices perpendicular with each other.   

In principle, increasing the virtual bond dimension $D$ (or Schmidt rank $\chi$ in 1D) can improve the results for any tensor network method. However, the choice of an ansatz following the geometric structure of the system can efficiently improve the calculation even at small $D$. With PESS on a checkerboard lattice, we obtain benchmark results of the ground state energies with small $D$ and better results can be obtained through either FU or further increasing $D$. For the results shown in the main text, we use $D=4$ with an $2\times 2$ unit cell ($n=2$ with eight local tensors) to compare the energies of different degenerate ground states (Fig.~2 in the main text) and use $D=4$ with a $\sqrt{2}\times\sqrt{2}$ unit cell ($n=1$ with four local tensors) to determine the phase transition boundaries (Fig.~3,4 in the main text). 

\section{Discussion on the nearly degenerate BCC/SCC-type states}
\subsection{Results with larger $D$}

\begin{figure}[h]
\includegraphics[width=0.7\textwidth]{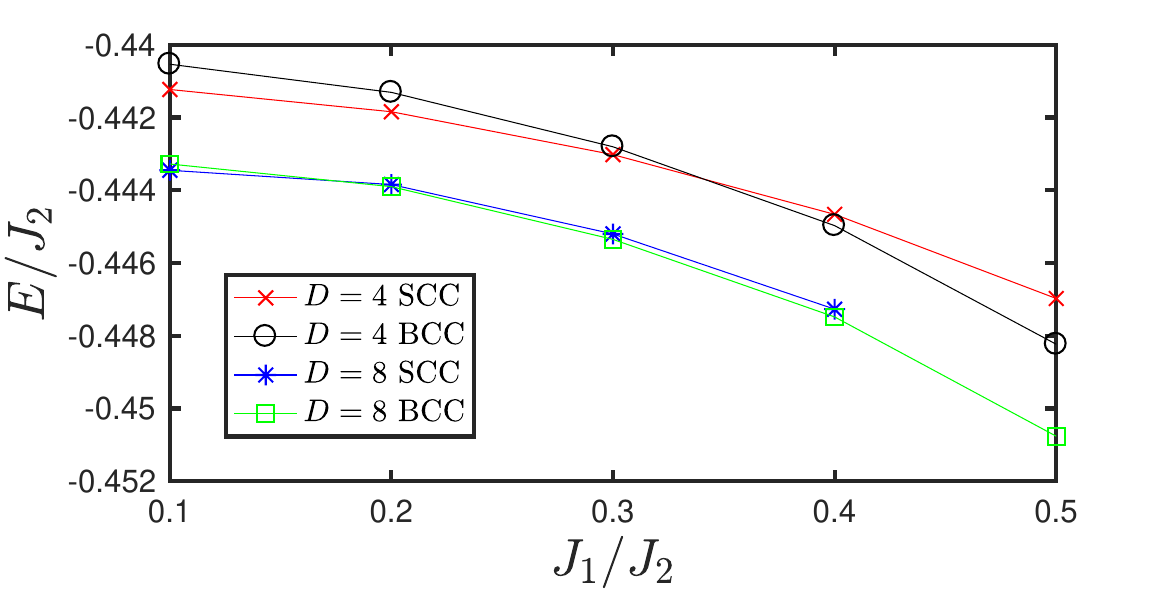}
\caption{Energy of the coexisted BCC and SCC state at $D=4$ and $D=8$ as functions of $J_1/J_2$. $D=8$ results give enhanced degenerate behavior with a weaker crossing at smaller $J_1$ (or larger $J_2$).}
\label{fig:largerD8}
\end{figure}
In the main text, we obtain nearly degenerate BCC and SCC states. Energies of these two kinds of states have a level crossing behavior from our $D=4$ calculation. This weak phase transition between the BCC and SCC states is also observed from larger $D$ calculation. At $D=5,6,$ and 7, we do not find stable BCC state at large $J_2/J_1$ or SCC states at small $J_2/J_1$, which is a direct evidence of one of our main results that the BCC (CD-VBS) is the ground state at $J_2\gtrsim J_1$ while the SCC states take over the rest at $J_2\gg J_1$. At $D=8$, where there is a similar coexisted region of the BCC and SCC states with the case in $D=4$, we find that the degeneracy between these two states are enhanced and the energy level crossing reappears at larger $J_2/J_1$.  In Fig.~\ref{fig:largerD8}, we compare the energies of the BCC state and one of the SCC states at $D=4$ and $D=8$ with a inverse $J_2$ unit ($J_2=1$) which can capture the level crossing at $10^{-4}$-$10^{-3}$ scale. Note that the data in the $J_2/J_1$ unit ($J_1=1$) in the main text are overlapped with each other. Thus, the nearly degenerate behavior and ground states arguments based on $D=4$ do consistent with larger $D$ calculation.   

\begin{figure}[h]
\includegraphics[width=0.7\textwidth]{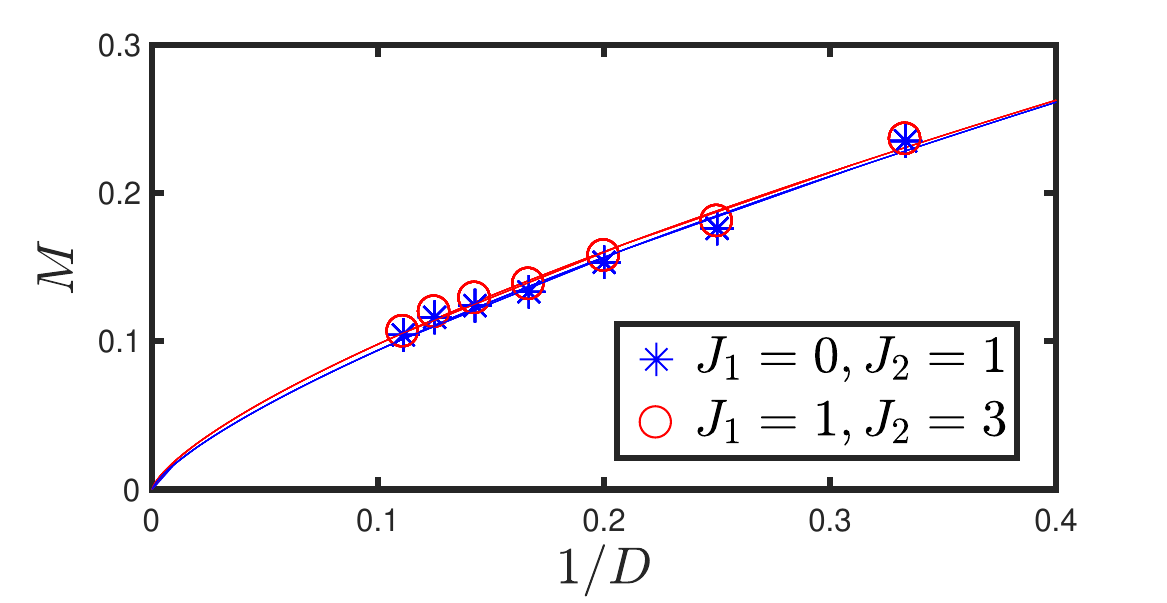}
\caption{Magnetization $M$ for $J_2/J_1=3$ and the 1D AF Heisenberg chain at $D=3,4,...$, and 9. Extrapolation with inverse $D$ shows that both cases give zero $M$ at infinite $D$. The lines are in a form $M=aD^{-b}$, with $a=0.5136,b=0.7375$ for 1D chain and $a=5037,b=7106$ for 2D case at $J_2/J_1=3$}
\label{fig:mag12}
\end{figure}

For the weakly coupled 1D AF SCC states, the different patterns are based on different local correlations along $J_1$ directions, not on the spin magnetic orders. To show this, we calculate the average magnetization on each site $M=\sqrt{\langle (S^x_i)^2\rangle+\langle (S^y_i)^2\rangle+\langle (S^z_i)^2\rangle}$ of the SCC state at finite $J_2/J_1$ and compare the results with those from the purely 1D Heisenberg chain at different $D$ [Fig.~\ref{fig:mag12}]. Two cases give the same extrapolating behavior with $D$ and suggest that a SCC state is a nonmagnetic ordered state.  

\subsection{Results with larger unit cell}

\begin{figure}[h]
\includegraphics[width=0.7\textwidth]{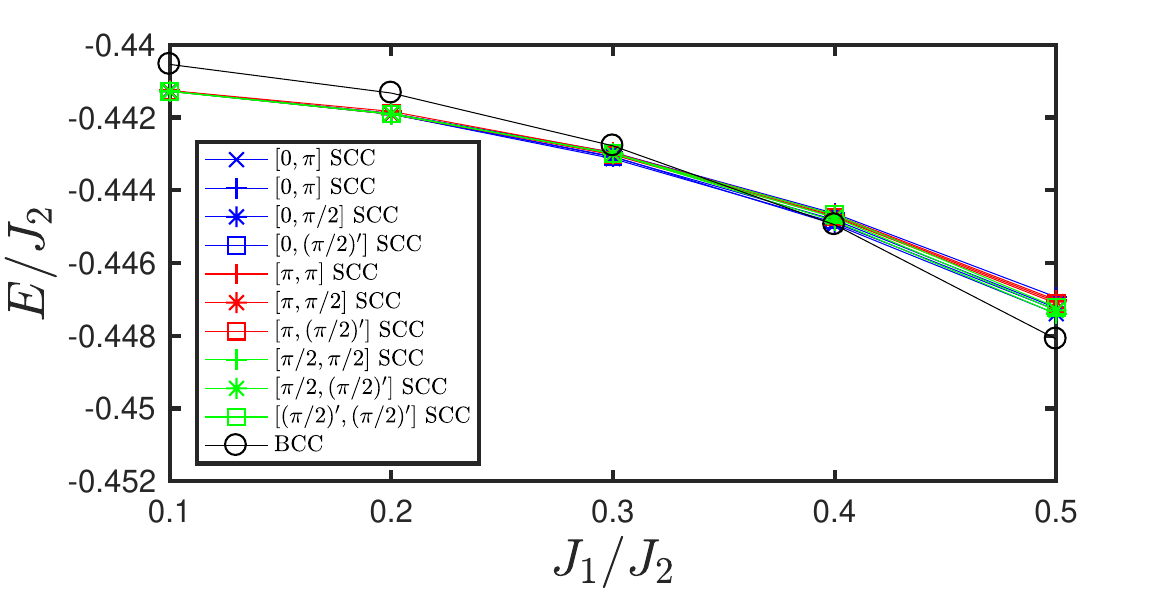}
\caption{Energies of 10 SCC states and the BCC state with functions of $J_1/J_2$. To classify the different SCC states, The notation $[x,y]$, $x=0,\pi,\pi/2$, and $(\pi/2)'$ is used, which is described in the main text and Sec.~\ref{notation}.}
\label{fig:largerUc}
\end{figure}

With a $2\times 2$ PESS unit cell, we obtain three SCC states with the ``stripe" state has a $\sqrt{2}\times\sqrt{2}$ structure, while the ``N\'eel*" and ``zigzag" states have a $2\times 2$ period. In fact, one can easily verify that these are the only SCC states within a $2\times 2$ unit cell. More nearly degenerate SCC states might be found with larger PESS unit cell. We push the unit cell limit to a $4\times 4$ structure with 32 local tensors inside and find more nearly degenerate SCC states, indeed. Figure~\ref{fig:largerUc} shows that there are 10 SCC states which have similar energies within a $4\times 4$ unit cell. In the $J_1/J_2$ unit ($J_2=1$), we can find clear energy level crossing of these 10 states with the BCC state, consistent with results based on a $2\times 2$ unit cell calculation shown in the main text. 

\subsection{Results in the $XY$-limit}
\begin{figure}[t]
\includegraphics[width=0.9\textwidth]{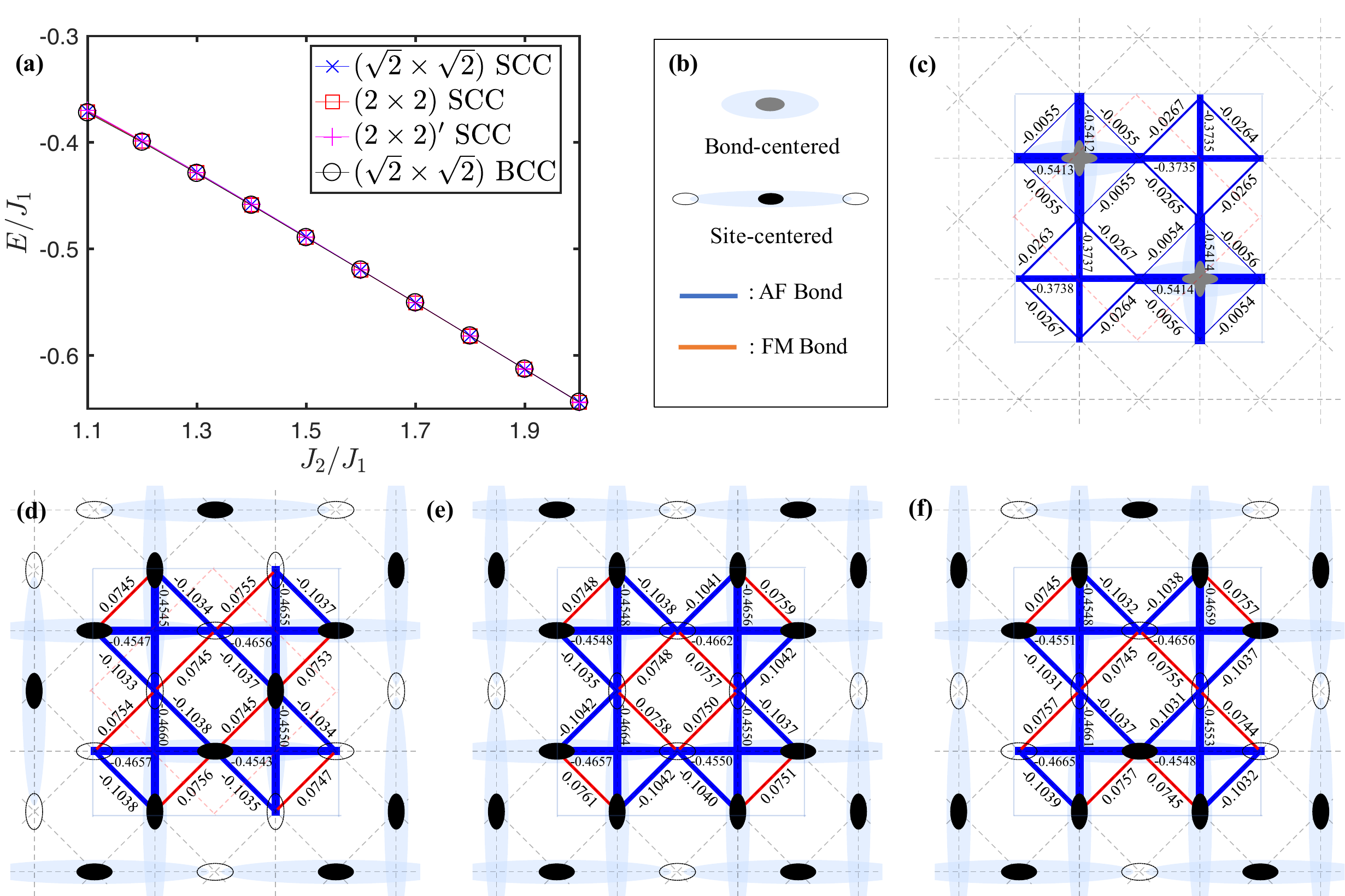}
\caption{Nearly degenerate ground state energies of checkerboard model at the $XY$-limit at $J_2>J_1$, notations are the same with those in Fig.~2 of the main text. (a) Energies of different states with different $J_2/J_1$. (b) Symbols used in (c-f) are illustrated. (c-f) Examples of the different energy bond configurations at $J_2/J_1=1.5$ are shown (a PESS unit cell is bounded by the light blue box): (c) $(\sqrt{2}\times\sqrt{2})$ BCC (crossed-dimer), (d) $(\sqrt{2}\times\sqrt{2})$ SCC (stripe), (e) $(2\times 2)$ SCC (N\'eel*), and (f) $(2\times 2)'$ SCC (zigzag). In (c,d), one period of the charge order is bounded by the dashed red lines. In (e,f), the period is the same with the PESS unit cell.}
\label{fig:xyconf}
\end{figure}

The nearly degenerating behavior of the BCC/SCC-type states found in the main text also exists in the $XY$-limit. As shown in Fig.~\ref{fig:xyconf} for the degenerate states in the $XY$-limit, there is no significant difference from the results in the Heisenberg case, except that the magnitude of the bond energies in $J_1$ direction are relatively larger than those in the Heisenberg case, which indicates that the local bosonic wave-function spreads further for the $XY$ case. This property is consistent with the fact that there is no density interaction in the $XY$-limit. 

\subsection{A new notation to classify the SCC states}
\label{notation}
\begin{figure}[h]
\includegraphics[width=0.8\textwidth]{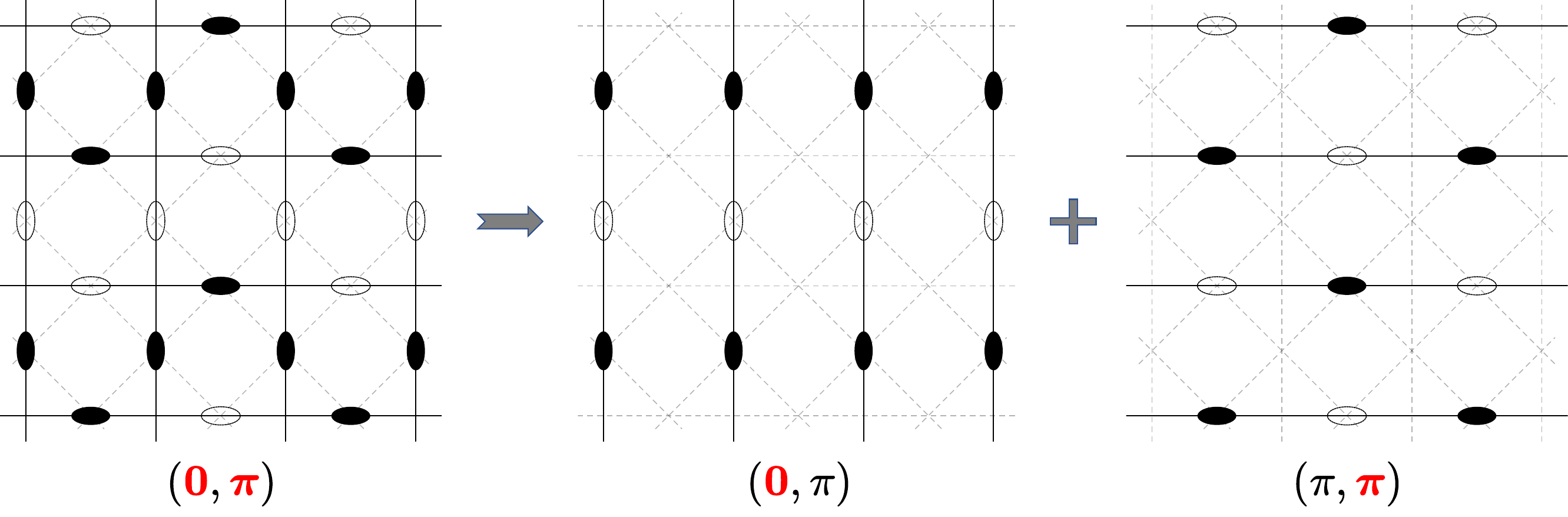}
\caption{A new notation to classify the SCC state for the zigzag state. The charge density patterns on a checkerboard lattice can be represented by the charge patterns in two different directions. In each direction, the charge is occupied at every other sites, labeled by a black $\pi$ in the nesting vector. The repeating pattern of the charge chain in the perpendicular direction is labeled by the red value. E.g., $0$ stands for the chain is repeated every lattice spacing.}
\label{fig:symbol}
\end{figure}

In the nearly degenerate BCC/SCC-type phase, both the N\'eel* and the zigzag correlated states have a $2\times 2$ structure. We used $(2\times 2)$ SCC and $(2\times 2)'$ SCC to distinguish these two SCC states. For the SCC states with a $4\times 4$ structure, another seven different correlated states appear. Here, we show that all the SCC states can be further classified by a combined nesting vector, i.e. $[x,y]\equiv [(x,\pi),(\pi,y)]$ ($\pi$ stands for the bosons occupied on every other lattice spacing), representing the repeating patterns of the vertical (horizontal) lines of bosons in the horizontal (vertical) directions. For example, the surprising zigzag state is labeled as $[0,\pi]$ (abbr. of $[(0,\pi),(\pi,\pi)]$) with the vertical lines of bosons repeat themselves every lattice spacing horizontally and horizontal lines repeat every other lattice spacing vertically (Fig.~\ref{fig:symbol}). Similarly, the stripe (N\'eel* state) is labeled by $[\pi,\pi]$ ([$0,0$]) and other seven states can be classified correspondingly. More degenerate states with all the possible combination of $[x,y]$ can be obtained if the size of the unit cell is further increased. 

\subsection{Discussion on the higher energy states from the bosonic picture}
We have shown that the nearly degenerate ground states can be interpreted by the bosonic picture. In all the SCC states, the bosons are occupied every other sites in the horizontal or vertical orbital directions (two $J_2$ bond directions). Meanwhile, there are two different types of $J_1$ bonds for each boson with equal probability: one is connect to a hole, the other is connect to another boson. In the BCC state, the crossed-dimer configuration is formed by two bosons in different orbital direction moving into the same plaquette with $J_2$ interactions. In this configuration, there are also two different types of $J_1$ bonds: one is the bonds inside the crossed-dimer plaquette, the other is those in between different crossed-dimer plaquette. Thus, the bond asymmetry is also appear in each $J_1$ direction. 

\begin{figure}[h]
\includegraphics[width=0.7\textwidth]{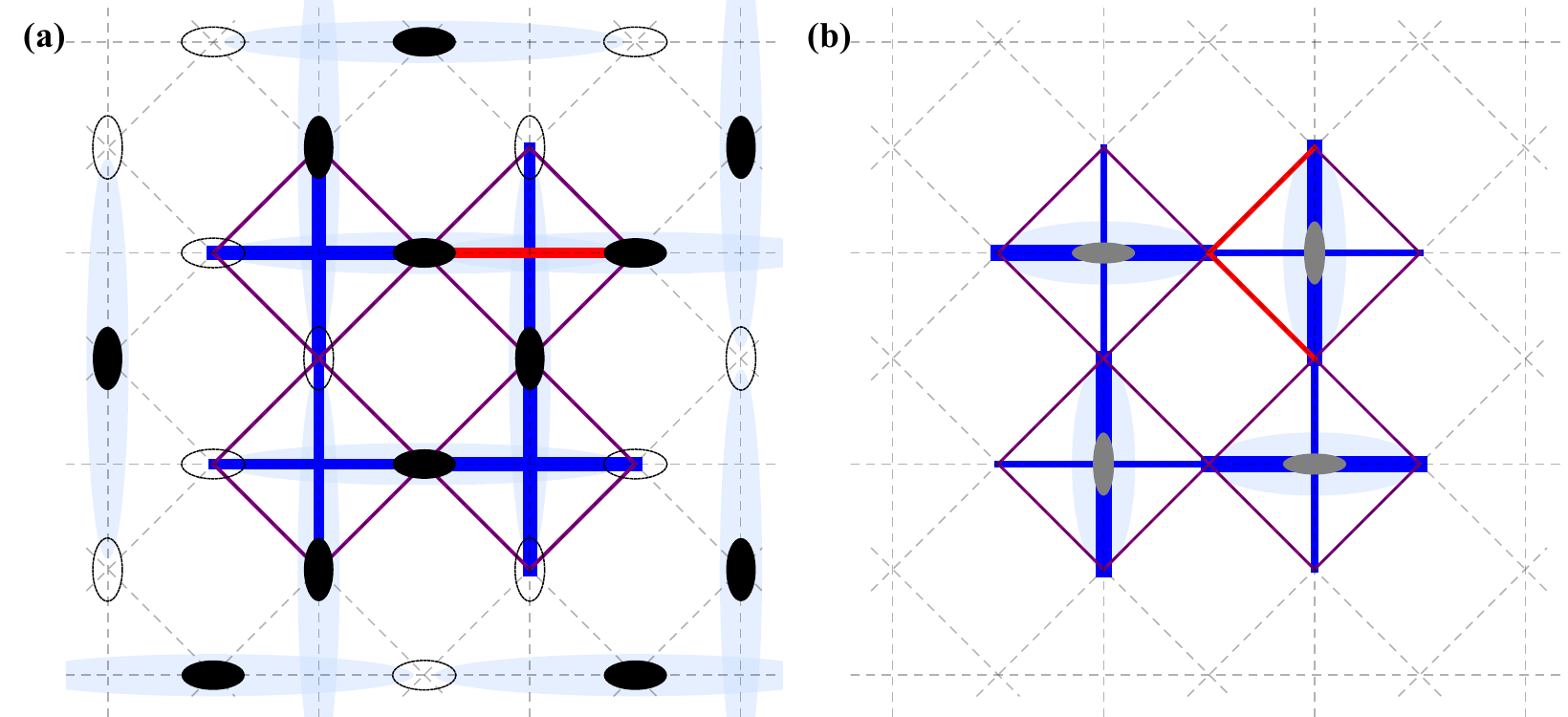}
\caption{Examples of higher energy states. (a) The configuration of a higher energy SCC state with two bosons are neighbors connected by a red line in the horizontal direction. (b) The SS configurations (a higher energy BCC state). The two red lines show the two nearest neighbor bonds shared by the same dimer.}
\label{fig:ss}
\end{figure}

To further illustrate the efficiency of the bosonic interpretation, we discuss the higher energy states. For a SCC state, moving a boson to a hole, either through $J_1$ or $J_2$ bond will generate a configuration with two bosons connected by a $J_2$ bond [Fig.~\ref{fig:ss}(a)], where the local wave-function of the two bosons overlap with each other, which result in a higher energy state. For the BCC case, beside the crossed-dimer state, a Shastry–Sutherland (SS) dimer configuration [Fig.~\ref{fig:ss}(b)] is another possible state where there is no overlap of bosonic wave-functions in the same $J_2$ orbital direction. This state is corresponding to the case when each boson is located at the center of each plaquette with $J_2$ interactions. However, different from the crossed-dimer BCC state where there are two different $J_1$ bonds, the $J_1$ bonds in the SS dimer state are all the same. They are all inter-dimer bonds and forming pairs of bonds connect to a common dimer [two red lines in Fig.~\ref{fig:ss}(b)], which also generate some overlap of the local wave-functions between bosons and result in a higher energy state. 

\begin{figure}[h]
\includegraphics[width=0.8\textwidth]{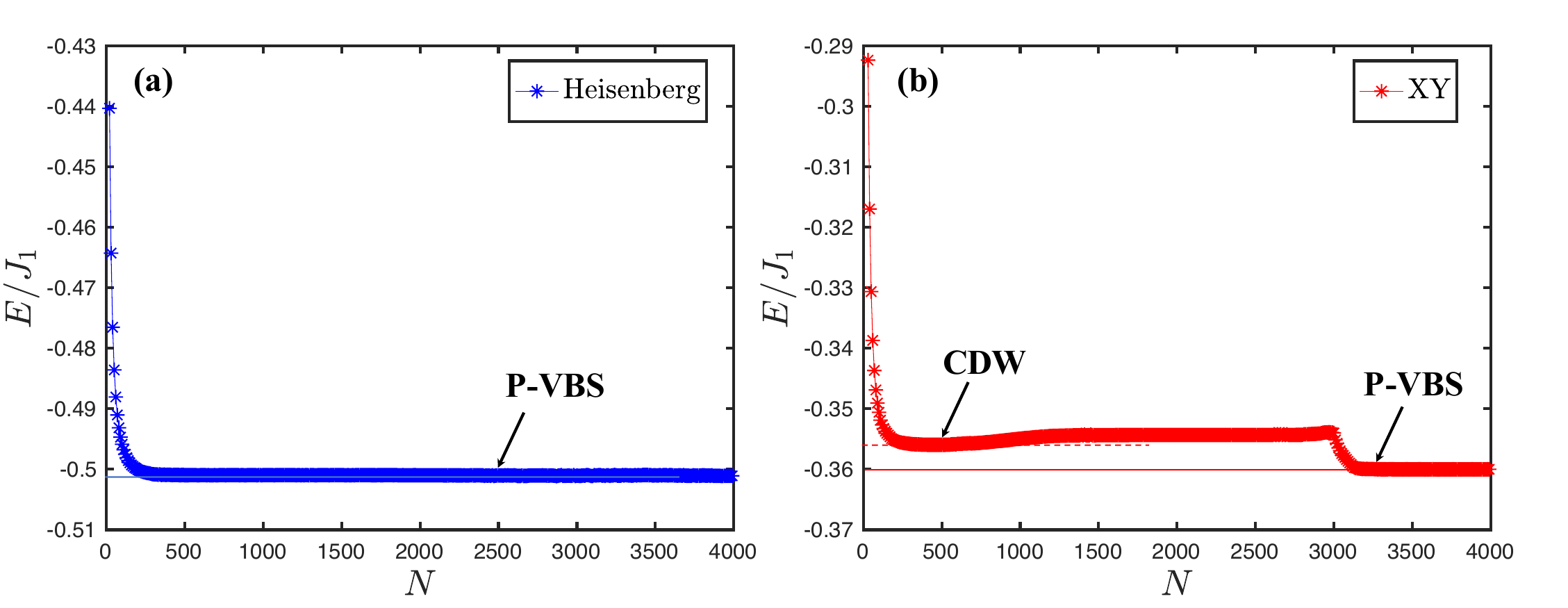}
\caption{The energy per site at $J_2/J_1=0.99$ as a function of the time evolution step (N) is shown. Starting from a classical N\'eel configuration, the system is evolved upto 4000 times and the energy is calculated at every ten steps. The energy ranges shown in both the Heisenberg and the $XY$ cases are taken as $0.08J_1$. (a) In the Heisenberg limit, the state goes monotonous to the ground state, a P-VBS state (thin blue line). (b) In the $XY$-limit, before ended at the P-VBS state (the solid thin red line), the state first evolves to a meta-stable CDW state (dashed thin red line).}
\label{fig:meta}
\end{figure}
\section{The meta-stable CDW state in the $XY$-limit}

At $J_2\sim J_1$, we find a meta-stable CDW state in the $XY$-limit. To illustrate the difference between the Heisenberg and the $XY$-limit, we start from a classical N\'eel configuration and evolve the state iteratively for both limits. Figure~\ref{fig:meta} shows the energy with a function of the evolution step $N$. For the Heisenberg limit, it goes monotonous to the P-VBS ground state. However, for the $XY$-limit, a meta-stable CDW state with a remaining AF ordered feature in the $z$ direction exist before it goes to the paramagnetic ground state. 

\begin{figure}[h]
\includegraphics[width=0.6\textwidth]{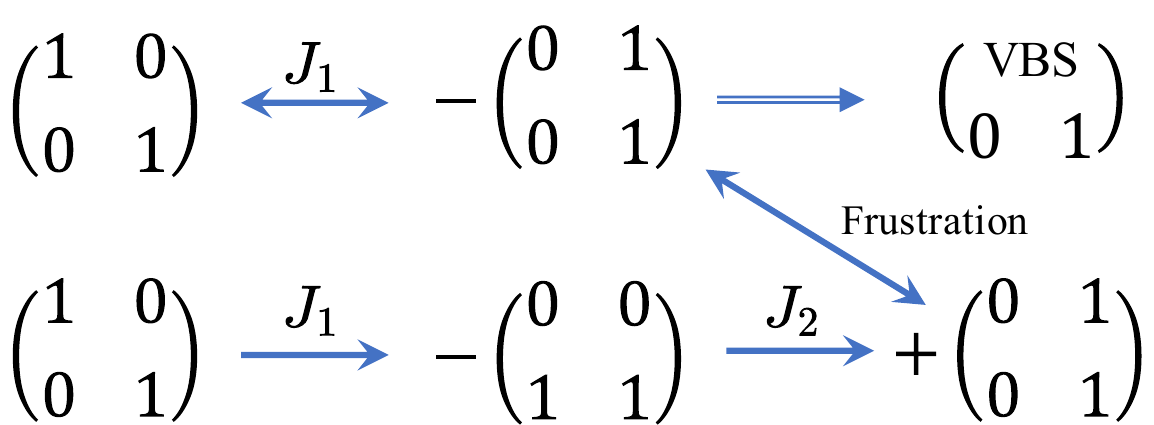}
\caption{The frustration due to the hopping at the $J_1$ and $J_2$ direction is illustrated in the configuration of one plaquette, where 1 and 0 stand for a a boson and a hole, respectively.}
\label{fig:cdw}
\end{figure}

The CDW state in the $XY$-limit can be understood in the bosonic picture. In the $J_1\gtrsim J_2$ region, the kinetic hopping along the $J_1$ and $J_2$ bond directions give opposite sizes of the same configuration with bosons occupied in the neighbor sites along $J_1$ directions in the wave-function (Fig.~\ref{fig:cdw}). This frustration behavior decouples two charge ordered states and breaks the symmetry of these two degenerate states to a CDW state.  

\begin{figure}[h]
\includegraphics[width=0.7\textwidth]{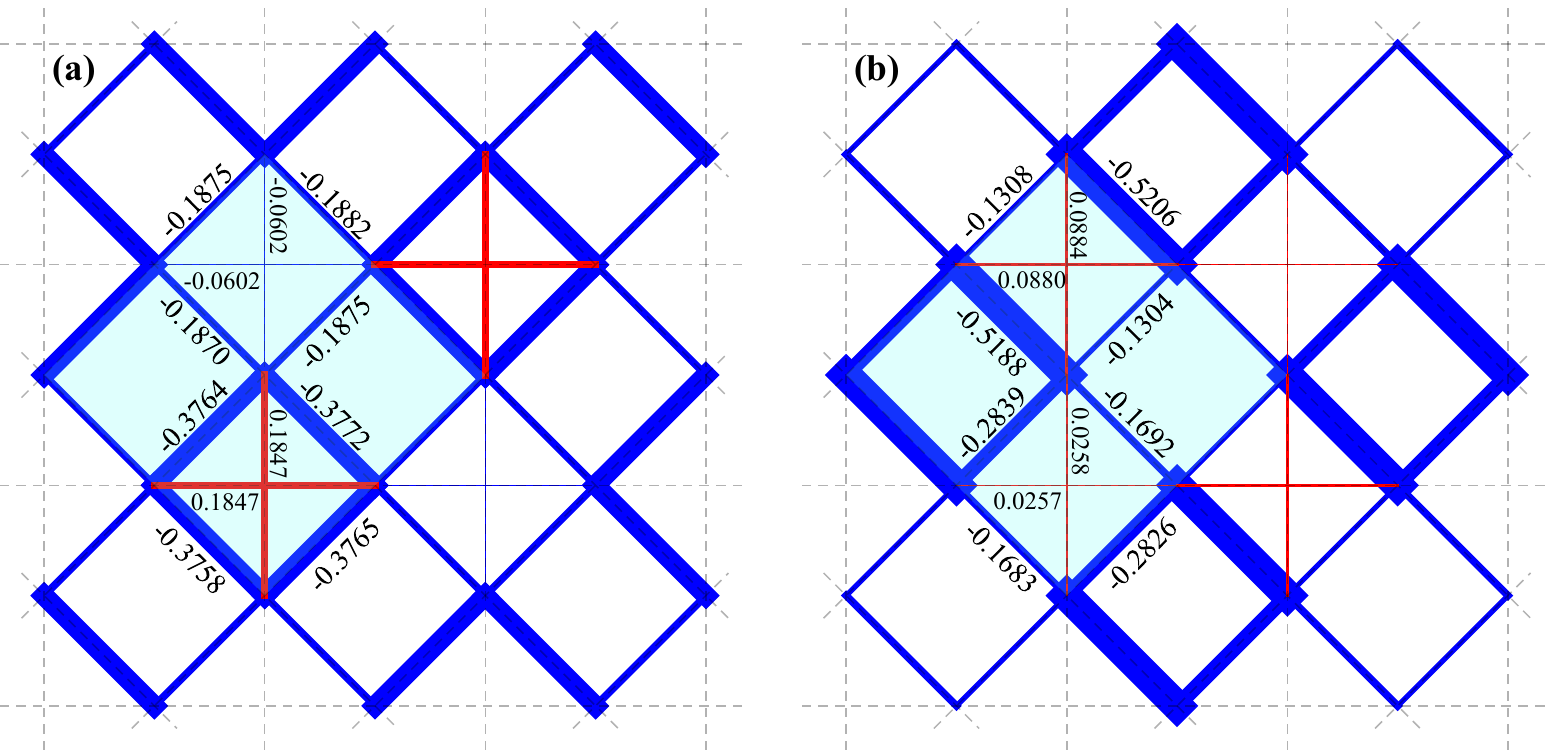}
\caption{Energy configurations for P-VBS states. (a) Example of the plaquettes on the squares with $J_2$ bonds at $J_1'/J_1=1$, $J_2/J_1=1$. (b) Example of the plaquettes on the squares without $J_2$ bonds at $J_1'/J_1=0.9$, $J_2/J_1=0.8$. The numbers show the bond energies and the thickness of each bond indicates the magnitude. The green regions represent a $\sqrt{2}\times\sqrt{2}$ PESS unit cell.}
\label{fig:pvbs}
\end{figure}

\section{Discussion on P-VBS states}
In the P-VBS state region, the plaquettes may sit either on the squares with or without $J_2$ bonds. The energies of both cases are lower than those of AF ordered configuration and the $J_2$-direction BCC/SCC-type configurations. In the isotropic limit ($J_1'/J_1=1$) and at $J_2\sim J_1$, the PESS calculation favors the plaquette configurations on the squares with $J_2$ bonds. Figure~\ref{fig:pvbs}(a) shows an example at $J_2=J_1$. However, in the wide region of the P-VBS state with the nematic perturbation $J_1'\neq J_1$, the PESS calculation favors the configuration without $J_2$ bonds in the plaquette. Figure~\ref{fig:pvbs}(b) shows an example at $J_1'/J_1=0.9$, $J_2/J_1=0.8$. The interplay between the two configurations inside the P-VBS state region deserves a future study with further modifying the PESS ansatz to fit both the plaquette and the $J_2$ crossing geometries. For example, constructing the simplex tensor $S$ on the plaquette without $J_2$ bonds and a four-virtual-leg local tensor $T$ in the PESS wave-function ansatz with a selected symmetric time evolution update scheme $H_p$ on all the plaquette (with and without the $J_2$ bonds) may provide a possible method to favor all the configurations of the P-VBS states. 

\begin{figure}[h]
\includegraphics[width=0.8\textwidth]{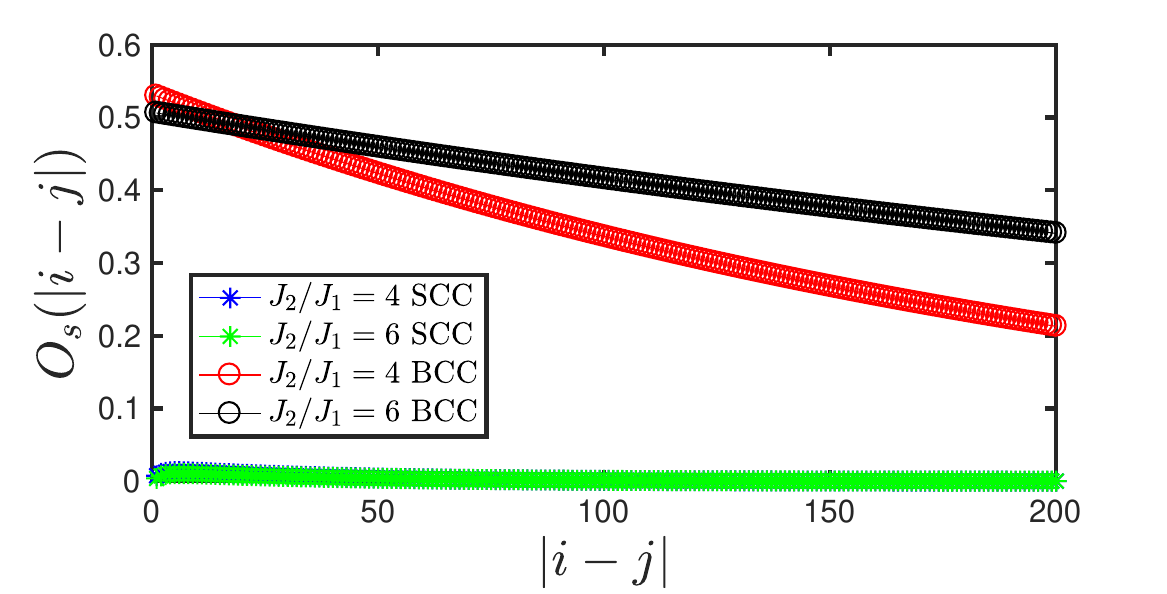}
\caption{String order parameter $O_s$ as functions of the distance at $J_2/J_1=4$ and 6 for both the SCC and BCC state. Slow decays of the $O_s$ of BCC state suggests a quasi-long ranged uudd correlation.}
\label{fig:string}
\end{figure}

\section{Discussion on quasi-long range uudd correlation}
Different experimental measurements on the pyrochlore material GeCu$_2$O$_4$ results in controversial results. For example, both 1D AF correlation~\cite{Yamada2000s} and up-up-down-down (uudd) type correlation~\cite{uudd2016s} are observed along the $J_2$ direction. Note that the GeCu$_2$O$_4$ is estimated to have $J_2/J_1\sim 6$, within the nearly degenerate region in our results. Besides that the SCC state gives a AF correlation along the $J_2$ direction, can the BCC state give insight to this material? On this basis, we calculate a string order parameter~\cite{string1989s,ZouSPT2019s,ZouSPT2020s} defined as
\begin{equation}
{O}_s(r)  =  -\! \lim_{r\to\infty}\langle(\hat{S}^z_{n} \! + \! \hat{S}^z_{n+1}) e^{ i \pi \! \sum_k \! \hat{S}^z_k }
(\hat{S}^z_{2r+n} \! + \! \hat{S}^z_{2r+n+1})\rangle,   
\end{equation}
for both the SCC and BCC states at different $J_2/J_1$,
where the $k$ sum is at $n+2\le k\le 2r+n-1$. A finite $O_s$ gives that the neighboring spins $\hat{S}^z_n+\hat{S}^z_{n+1}$ form an effective spin-1 degree of freedom and suggests an uudd type correlation. As shown in Fig.~\ref{fig:string}, we find that for the SCC state where there is a AF correlation, $O_s(r)$ decay very fast to zero as $r$ increase as expected. However, for the BCC state, a surprise large $O_s(r)$ which decay very slowly and shows a quasi-long ranged uudd type of correlation. As $J_2/J_1$ increases, the uudd correlation is enhanced. This string order parameter result shows that the GeCu$_2$O$_4$ can be settled in a BCC state and the near degeneracy of the SCC/BCC states implies that the system is highly susceptible to various weak perturbations which may results in different symmetry breaking states. 

%

\end{document}